\documentclass[%
 table,
 superscriptaddress,
 amsmath,amssymb,
 aps,
 pra,
]{revtex4-2}

\usepackage[utf8]{inputenc}
\usepackage[T1]{fontenc}
\usepackage{amssymb}
\usepackage{mathtools}
\usepackage{bbold}
\usepackage{bm}
\usepackage[position=top, caption=false]{subfig}
\usepackage[table,x11names,dvipsnames,table]{xcolor}

\usepackage{amsthm}
\usepackage{amsmath}
\theoremstyle{definition}

\DeclareMathOperator*{\argmin}{arg\,min}

\begin{document}


\title{Whole-Device Entanglement in a 65-Qubit Superconducting Quantum Computer}



\affiliation{School of Physics, University of Melbourne, VIC, Parkville, 3010, Australia.}
\affiliation{School of Mathematics and Statistics, University of Melbourne, VIC, Parkville, 3010, Australia.}

\author{Gary J Mooney}
\email{mooneyg@unimelb.edu.au}
\affiliation{School of Physics, University of Melbourne, VIC, Parkville, 3010, Australia.}
\author{Gregory A L White}
\email{white.g@unimelb.edu.au}
\affiliation{School of Physics, University of Melbourne, VIC, Parkville, 3010, Australia.}
\author{Charles D Hill}
\email{cdhill@unimelb.edu.au}
\affiliation{School of Physics, University of Melbourne, VIC, Parkville, 3010, Australia.}
\affiliation{School of Mathematics and Statistics, University of Melbourne, VIC, Parkville, 3010, Australia.}
\author{Lloyd C L Hollenberg}
\email{lloydch@unimelb.edu.au}
\affiliation{School of Physics, University of Melbourne, VIC, Parkville, 3010, Australia.}

\date{\today}



\begin{abstract}

The ability to generate large-scale entanglement is an important progenitor of quantum information processing capability in noisy intermediate-scale quantum (NISQ) devices. In this paper, we investigate the extent to which entangled quantum states over large numbers of qubits can be prepared on current superconducting quantum devices. We prepared native-graph states on the IBM Quantum 65-qubit \textit{ibmq\_manhattan} device and the 53-qubit \textit{ibmq\_rochester} device and applied quantum readout-error mitigation (QREM). Connected entanglement graphs spanning each of the full devices were detected, indicating bipartite entanglement over the whole of each device. The application of QREM was shown to increase the observed entanglement within all measurements, in particular, the detected number of entangled pairs of qubits found within \textit{ibmq\_rochester} increased from 31 to 56 of the total 58 connected pairs. The results of this work indicate full bipartite entanglement in two of the largest superconducting devices to date.

\end{abstract}

\maketitle


\section{Introduction} \label{sec:introduction}
In the context of quantum computing, entanglement is usually characterized by the presence of non-classical correlations between qubits~\cite{horodecki2009quantum}. It can be thought of as a resource unique to quantum computation~\cite{wootters1998quantum, shor1999polynomial, raussendorf2001one} that is at the heart of achieving quantum speedup~\cite{preskill2012quantum, vidal2003efficient, verstraete2004renormalization, verstraete2008matrix} as well as playing an important role in fundamental physics -- being the phenomenon behind Einstein's ``spooky action at a distance''~\cite{einstein1935can}. The development of noisy intermediate-scale quantum (NISQ) devices~\cite{preskill2018quantum} is rapidly advancing, with ``quantum supremacy'' demonstrated on sampling problems~\cite{arute2019quantum, zhong2020quantum}. To measure the advance of quantum computer technology, a variety of metrics within the quantum computing literature exist that encapsulate progress in a range of categories such as qubit coherence times, cross-talk, state preparation and measurement, and fidelities of gate operations \cite{flammia2011direct,da2011practical,cross2019validating,sarovar2020detecting,Blume-Kohout2017,Proctor2019DRB}.
As the field progresses towards larger multi-qubit systems, an appropriate metric to summarize the performance of a quantum information processor is the generation and characterization of large entangled states.
Such states can be generated in the form of either mixed state bipartite entanglement, where the state is inseparable with respect to all bipartitions of qubits within the system, or the stronger condition of genuine multipartite entanglement (GME), where the state always contains inseparable pure states spanning all of its qubits.
These two forms of entanglement are described in more detail in~\cite{mooney2019entanglement}. Alternatively, the detailed structure of entanglement can be determined for a deeper analysis~\cite{lu2018entanglement}. For systems with full qubit control, large states that have been demonstrated to be fully bipartite entangled in recent years consist of 16 and 20 qubits in superconducting systems~\cite{wang201816,mooney2019entanglement}, and 20 qubits in ion trap systems~\cite{friis2018observation}. The stronger condition of genuine multipartite entanglement has recently been shown on systems of sizes 18 and 27 qubits in superconducting systems~\cite{wei2020verifying, mooney2021generation}, 12 photons in a photonic system (polarization)~\cite{zhong201812}, 18 qubits in a photonic system (multiple degrees of freedom)~\cite{wang201818}, 20 qubits in a neutral atom system~\cite{omran2019generation}, and 24 qubits in an ion trap system~\cite{pogorelov2021compact}. There has also been recent work showing violations of robust Bell inequalities for path graph states on up to 57 qubits on the 65-qubit superconducting \textit{ibmq\_manhattan} device~\cite{yang2021testing}.

A variety of superconducting quantum devices are hosted by IBM Quantum's cloud service~\cite{ibmq}, ranging in sizes from 1 to 65 qubits. Extensive reviews on superconducting qubits are provided in the works~\cite{kjaergaard2020superconducting, Kockum2019, gu2017microwave}. The two largest of the IBM Quantum devices are the \textit{ibmq\_rochester} and \textit{ibmq\_manhattan} devices, consisting of 53 and 65 qubits respectively. Although \textit{ibmq\_rochester} has a large number of qubits, the error rates are relatively high in comparison to other IBM Quantum devices, with an average readout error rate of $\sim$ 13\% and coherence times of T$_1$, T$_2$ $\sim$ 53 $\mu$s~\cite{ibmq}. The \textit{ibmq\_manhattan} on the other hand has relatively low error with an average readout error rate of $\sim$ 2\% and coherence times of T$_1$ $\sim$ 60 $\mu$s and T$_2$ $\sim$ 78 $\mu$s~\cite{ibmq}. In this work, we investigate the extent to which whole-device entangled quantum states can be prepared on IBM Quantum devices such as \textit{ibmq\_rochester} and \textit{ibmq\_manhattan}. A native-graph state (which has an edge corresponding to each connected pair of qubits within the device) was prepared on the devices, and quantum state tomography was performed on each pair of connected qubits and their neighbors. Using these measurements, the negativity~\cite{plbnio2007introduction, christandl2006structure} was calculated to detect entanglement between each pair of connected qubits. The measurement process of reading out physical qubits is imperfect, leading to a significant amount of readout error that can obfuscate the resulting data. This can cause prepared quantum states within the device to appear much less entangled than they actually are. To address this issue, a quantum readout-error mitigation (QREM) technique~\cite{maciejewski2020mitigation} based on quantum detector tomography (QDT)~\cite{lundeen2009tomography}, that is commonly used on superconducting qubits~\cite{neeley2010generation,dicarlo2010preparation,barends2014superconducting,song201710, wei2020verifying,mooney2021generation} is used to correct raw data. For both devices, we find that the entangled pairs of qubits form connected entangled graphs that include all qubits of the devices, indicating the states were fully bipartite entangled. The results for the \textit{ibmq\_manhattan} device show clear whole-device entanglement, whether QREM was used or not, with all connected pairs of qubits exhibiting entanglement. However, for the \textit{ibmq\_rochester} device, our results indicate that the application of QREM was crucial to mitigate the relatively large readout errors and observe whole-device entanglement, with the number of entangled pairs of qubits increasing from 31 to 56 from the total 58 connected pairs.

\begin{figure}
\centering
   \includegraphics[scale=1]{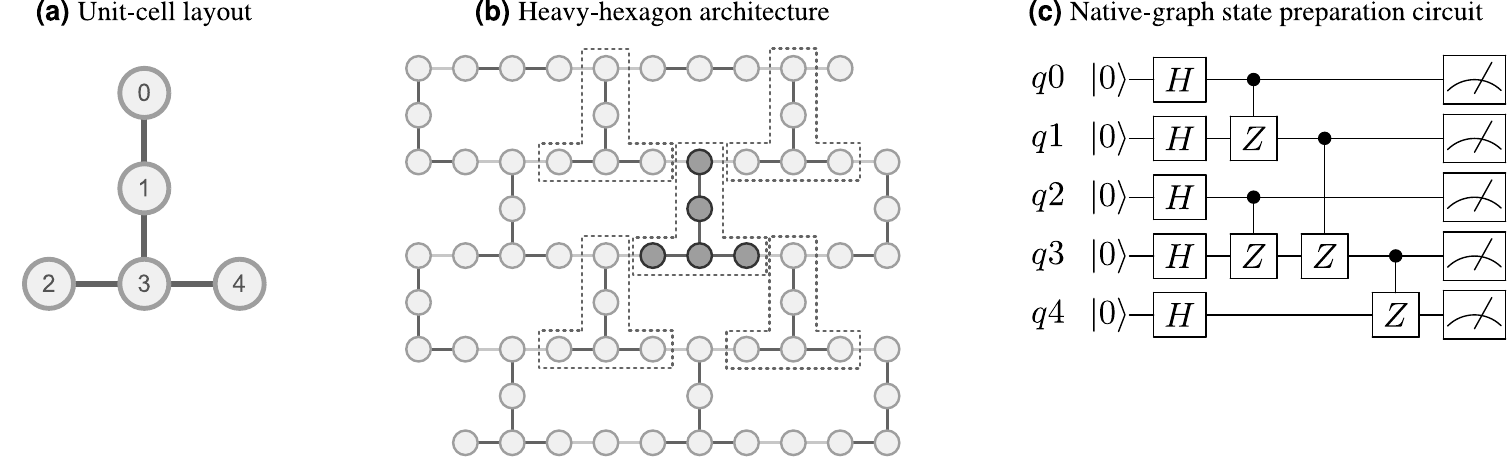}
	\caption{An example of preparing a native-graph state on a heavy-hexagon unit-cell layout. \textbf{(a)} A heavy-hexagon unit-cell layout with q3 being adjacent to q1, q2 and q4. The unit cell is defined to help visualize how the native-graph state is prepared across the full devices with a CNOT circuit depth of only three. \textbf{(b)} The heavy-hexagon architecture for the \textit{ibmq\_manhattan} device. The graphic shows how the unit cells can be patterned across the layout, where they are stitched together with controlled-phase gates to prepare the native-graph state over the full device. \textbf{(c)} The circuit to prepare the graph state for the heavy-hexagon unit cell. There are three controlled-phase gates applied to q3, leading to a CNOT depth of three for the circuit. To generate the native-graph state for the full device, the unit cells can be stitched together by using additional controlled-phase gates without needing to increase the circuit depth. For example, three unit cells $U_0$, $U_1$ and $U_2$ can be stitched together by applying two controlled-phase gates, one from q0 in $U_1$ to q4 in $U_2$ during the second layer of controlled-phase gates and one from q0 in $U_1$ to q2 in $U_0$ during the third layer. } \label{fig:graph-state-prep-example}
\end{figure}

\section{Results}

\subsection{Generating and detecting entanglement in graph states}
Graph states are entangled states that are defined in relation to a particular connected graph~\cite{hein2004multiparty}. Vertices represent qubits that are prepared in the $|+\rangle := (|0\rangle + |1\rangle)/\sqrt{2}$ state by applying a Hadamard gate to their initial $|0\rangle$ state and edges represent pairs of qubits that are acted upon by controlled-phase gates. A graph state can be expressed as
    \begin{equation}
        |G_N\rangle = \prod \limits_{(\alpha,\beta)\in E} \mathrm{CZ}^\alpha_{\beta} \;|+\rangle^{\otimes N},
    \end{equation}
where $E$ is the edge set of the graph $G_N$ corresponding to the $N$-qubit graph state, and $\mathrm{CZ}^\alpha_{\beta}$ represents a controlled-phase gate between connected qubits~$\alpha$ and~$\beta$. Graph states have a distinct advantage over other highly entangled states because they can be prepared in parallel, allowing them to be prepared in constant circuit depth with respect to qubit count. Additionally, they have been shown to have entanglement that is more robust against local measurements and noise than Greenberger-Horne-Zeilinger (GHZ) states prepared over the same qubits~\cite{briegel2001persistent}. Preparing a large graph state is related to showing the power of one-way quantum computation~\cite{raussendorf2001one,tanamoto2006producing,you2007efficient,tanamoto2009efficient}. In one-way quantum computing, all of the entanglement is created in the initialization of the state, then any quantum computation can be performed using only single-qubit measurement operations. To measure the extent of entanglement within a quantum device, we attempt to prepare a native-graph state which contains all qubits of the device by having an edge for every connected pair of qubits. An example of a native-graph state preparation circuit is shown in Figure~\ref{fig:graph-state-prep-example}. The CNOT depth of the preparation circuit is at least as long as the maximum number of neighboring qubits any single qubit is adjacent to within the device. This is because a controlled-phase gate is applied between the single qubit and each of its neighbors.

\begin{figure}
	\centering
    \includegraphics[scale=1]{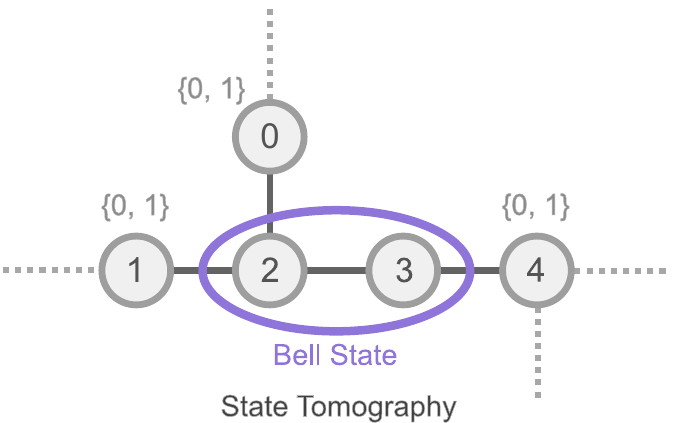}
	\caption{An example of measuring the negativity between arbitrary adjacent qubits q2 and q3 from an experimentally prepared native-graph state on a heavy-hexagon layout. Full quantum state tomography is performed on the pair and their nearest neighbors, that is, qubits q0, q1, q2, q3 and q4. The neighboring qubits q0, q1 and q4 are projected onto the $Z$-basis states 0 and 1 for each possible combination, that is, 000, 001, 010, 011, 100, 101, 110, 111. The negativity is measured for all combinations and the highest negativity is chosen to represent the extent of entanglement between the two qubits. The neighboring qubits are included in tomography and manually projected in this way because it is important to control which state the qubits are projected onto, since different combinations of projected $Z$-basis states may produce different Bell states. Thus tracing out the neighboring qubits without projecting them onto particular states may cause the remaining 2-qubit state to look maximally mixed.} \label{fig:tomography-and-negativity-example}
\end{figure}

A mixed state $\rho$ is separable if it can be expressed as a probabilistic mixture of separable pure states with respect to fixed qubit bipartitions $A$ and $B$, that is,
\begin{equation}
    \rho = \sum_i^N p_i \rho_i^{A} \otimes \rho_i^{B},
\end{equation}
where $\rho_i^A$ and $\rho_i^B$ are pure states of $A$ and $B$ respectively, $N$ is the number of pure states over all qubits in the composition, and the probabilities satisfy $p_i\geq 0$ and $\sum_i p_i = 1$. A mixed state is bipartite entangled if it is not separable with respect to all bipartitions of qubits within the system. To detect bipartite entanglement, instead of using full quantum state tomography on the entire quantum state (which infeasibly scales as $3^N$ circuits, since there are four basis states for each qubit where measurements that include the~$I$ basis can be obtained through post processing) we show that entanglement is present between partitions of any bipartition of qubits~\cite{wang201816,mooney2019entanglement}. By measuring the entanglement between every adjacent pair of qubits in the quantum state, an entangled graph can be constructed with edges present only when the pair is found to be entangled. Connected regions of entangled graphs represent entangled clusters of qubits, since they cannot be bipartitioned into disconnected (separable) partitions. Thus, generating a connected entangled graph spanning the full graph state is equivalent to showing that the state is inseparable with respect to any fixed bipartition of qubits and is therefore fully bipartite entangled. The graph representation of the native-graph state for any particular system naturally maximizes the number of possible cycles, enabling the corresponding prepared state to tolerate the highest number of non-entangled pairs while still having a connected entangled graph that spans all qubits. Graph states are particularly convenient for measuring the entanglement between adjacent pairs of qubits. They have the property that by projecting all but two qubits of the state, the remaining two qubits can be transformed to a Bell state (up to local transformations)~\cite{raussendorf2003measurement}. This is related to the generation of a Bell state through one-way quantum computation, where the logical qubits are adjacent to one another in the graph state. Entanglement can be measured between the pair of qubits by measuring the negativity~\cite{plbnio2007introduction, christandl2006structure}. For a quantum state represented as a density matrix $\rho$, the negativity $\mathcal{N}(\rho)$ between qubit bipartitions $A$ and $B$ can be calculated as
\begin{equation}
    \mathcal{N}(\rho) := \sum\limits_i \frac{|\lambda_i| - \lambda_i}{2},
\end{equation}
where $\lambda_i$ are the eigenvalues of the partial transpose of~$\rho$ with respect to bipartition~$B$~\cite{vidal2002computable}. Two qubits are entangled if and only if the negativity between them is non-zero~\cite{peres1996separability, horodecki2001separability}. To experimentally observe this, quantum state tomography is performed for each adjacent pair of qubits and their neighboring qubits within the graph state, as shown in Figure~\ref{fig:tomography-and-negativity-example}. Due to imperfections in measurement, the resulting density matrix may have negative eigenvalues, making it non-physical. The nearest physical density matrix under the 2-norm is obtained using an efficient algorithm by Smolin et al.~\cite{smolin2012efficient}. The neighboring qubits are then projected onto each combination of $Z$-basis states to obtain Bell pairs (in principle). These pairs are used to calculate negativities and the largest negativity among the combinations of $Z$-basis state projections is used to indicate the extent of entanglement within the pair.

\subsection{Quantum readout-error mitigation (QREM)}

\begin{figure}[!htbp]
	\centering
    \includegraphics[scale=1]{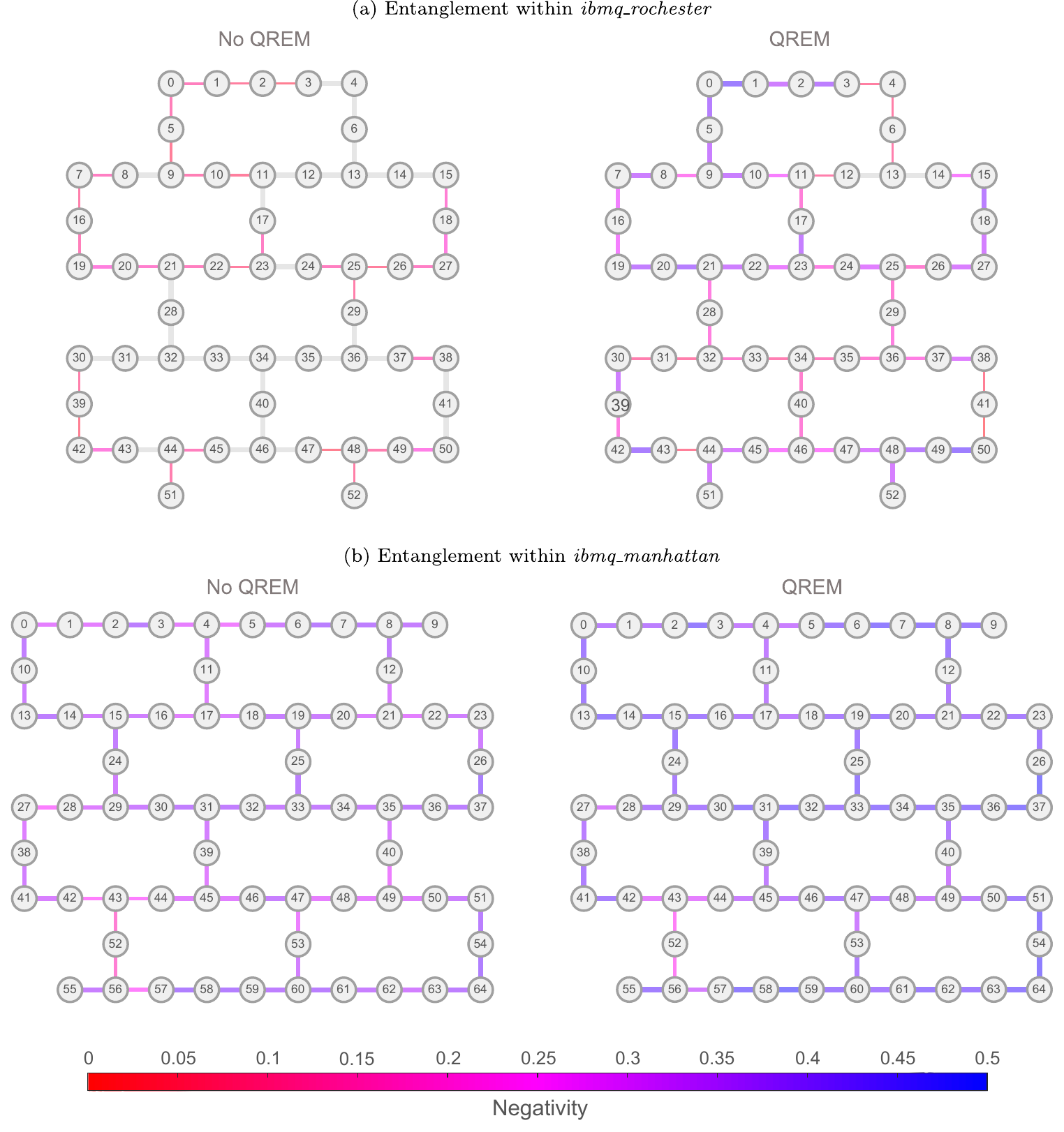}
	\caption{A graphic visualizing the entangled pairs of qubits detected within native-graph states prepared on the 53-qubit \textit{ibmq\_rochester} and 65-qubit \textit{ibmq\_manhattan} devices. Native-graph states are graph states spanning all qubits of the device with entangling gates performed on each connected pair of qubits in the layout. They can be prepared with a CNOT circuit depth of three using the strategy shown in Figure~\ref{fig:graph-state-prep-example}, where heavy-hexagon unit cells are prepared in parallel and stitched together using controlled-phase gates. Colored lines indicate that entanglement was detected between the corresponding pair of qubits with 95\% confidence and light gray lines indicate that entanglement was not detected with confidence. Thick blue lines indicate high negativity maximizing at 0.5, which corresponds to maximal entanglement. Conversely, thin red lines indicate low negativity minimizing at 0, which corresponds to no entanglement (in which case the edge is colored light gray). Connected subgraphs induced by entangled pairs of qubits are considered to be entangled regions of the device. \textbf{(a)} Entanglement within the \textit{ibmq\_rochester} device. When quantum readout-error mitigation (QREM) was not applied (left), 31 of the 58 pairs were found to be entangled with the largest entangled region consisting of 9 qubits. When QREM was applied (right), a total of 56 out of 58 connected pairs of qubits were found to be entangled with the entangled region consisting of all qubits of the device. The corresponding negativity values used to determine entanglement are shown in Figure~\ref{fig:negativity-plots}a. \textbf{(b)} Entanglement within the \textit{ibmq\_manhattan} device. In both cases of applying and not applying QREM, all 72 pairs of connected qubits were found to be entangled. Thus, the entangled region includes all qubits of the device. The corresponding negativity values are shown in Figure~\ref{fig:negativity-plots}b.} \label{fig:entanglement-layouts}
\end{figure}

Due to error introduced during the measurement process, quantum states are often significantly more entangled than indicated by entanglement measures. Even highly entangled states can have measurements fail to detect entanglement. 
This breakdown can stem from physical errors which may be either classical or quantum in nature, or to a lesser extent, statistical noise due to the limited number of shots. For IBM Quantum devices, it has been shown that the dominant form of measurement noise is classical, motivating a QREM technique used to alleviate measurement error~\cite{maciejewski2020mitigation}. The measurement noise is represented as a stochastic calibration matrix $\Lambda$ that contains the conditional probabilities for measuring each erroneous state given each ideally measured state. The calibration matrix is constructed using QDT~\cite{lundeen2009tomography} where combinations of all measurement states in the computational basis are used as the basis states. To build a calibration matrix that considers readout-error correlations between all qubits, the number of basis state measurements required is $2^N$ where $N$ is the number of qubits. For large quantum devices, the number of circuits is too impractical to build the full calibration matrix using QDT. Instead, we assume that the errors are uncorrelated. This allows the calibration matrix, $\Lambda$, to be written as
\begin{equation}
    \Lambda := \bigotimes_{i=1}^n \Lambda_i
\end{equation}
where $\Lambda_i$ is the calibration matrix for an individual qubit $i$ in the computational basis, defined as 
\begin{equation}
\label{cal-mat}
    \Lambda_i := \begin{pmatrix} p_i(0|0) & p_i(0|1) \\ p_i(1|0) & p_i(1|1) \end{pmatrix}, 
\end{equation}
where the notation $p_i(x|y)$ indicates the probability of measuring qubit $i$ in the state $|x\rangle$ given the prepared state $|y\rangle$, where $|x\rangle$ and $|y\rangle$ are in the computational basis. This simplification reduces the circuit count to a constant of two since the basis states for each of the qubits can be measured in parallel. In our previous work~\cite{mooney2021generation}, the assumptions made in performing QREM in this way on a similar IBM Quantum device (27-qubit \textit{ibmq\_montreal}) were investigated in detail. These include the assumptions that measurement error is significantly larger than preparation error, that measurement error is predominantly classical, and that measurement error is mostly uncorrelated between qubits. It was shown on sets of four qubits that there were only small differences between full calibration matrices and ones that assume uncorrelated measurement error. This is consistent with another work~\cite{maciejewski2020mitigation} that showed correlations between readout errors of pairs of qubits on the IBM Quantum's five-qubit \textit{ibmqx4} device are typically small in most but not all cases. Once the calibration matrix $\Lambda$ has been constructed, it can be inverted and applied to the measured probability vectors $\bm{p}_\text{exp}$ obtained from tomography measurements before the state is constructed as a density matrix, to correct for classical noise. However, the resulting vector is not always physical and may contain negative probabilities due to other forms of noise. To overcome this, the closest physical probability vector $\bm{p^{*}}$ to $\Lambda^{-1}\bm{p}_\text{exp}$ under the Euclidean norm can be found using the equation
\begin{equation}
    \bm{p^{*}} = \argmin_{\{\bm{p^{\prime}}\; |\; \forall i\; p^\prime_i \geq 0,\; \sum_{i=1}^n p^\prime_i = 1\}}\left(||\Lambda^{-1} \bm{p}_\text{exp} - \bm{p^{\prime}}||_2\right).
\end{equation}
This can be solved by using an efficient algorithm that finds the maximum-likelihood probability distribution to a set of real values summing to one~\cite{smolin2012efficient}.

\subsection{Entanglement of graph states in the \textit{ibmq\_rochester} and \textit{ibmq\_manhattan} devices}

\begin{figure}
	\centering
    \includegraphics[scale=1]{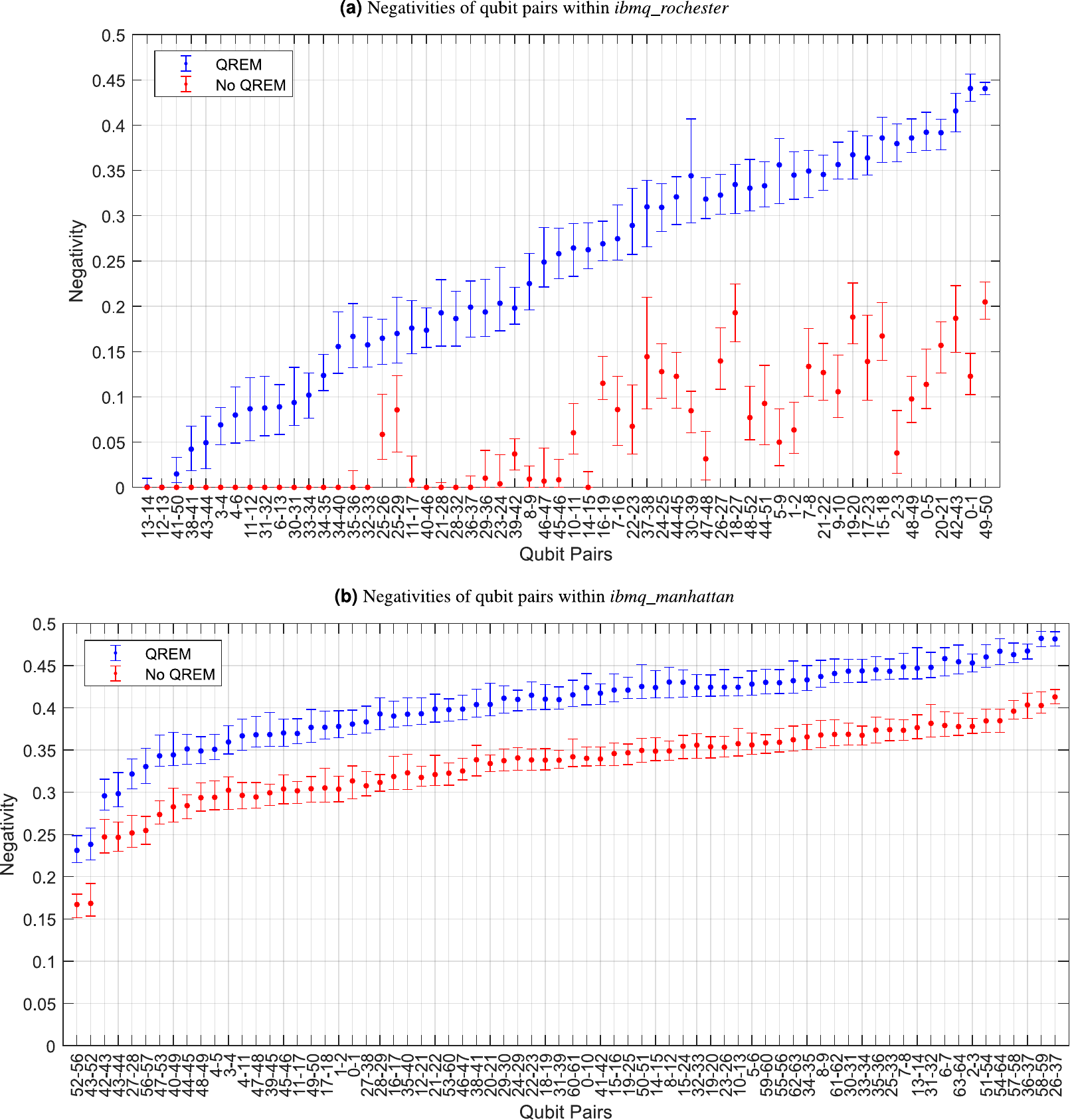}
	\caption{Negativity experimental results on the IBM Quantum devices showing 53-qubit and 65-qubit entanglement. Each adjacent pair of qubits $i$ and $j$ of the native-graph state is represented as $i$-$j$ in the figure. The plotted data are the highest negativity values among combinations of neighboring qubit $Z$-basis projections between each adjacent pair of qubits. The negativity ranges between 0 and 0.5 where 0 indicates no entanglement and 0.5 indicates maximal entanglement. The 95\% confidence intervals are calculated using bootstrapping methods~\cite{efron1986bootstrap, efron1981nonparametric}. The data are arranged in ascending order of confidence interval lower bounds in the case of QREM. \textbf{(a)} The negativity results for the \textit{ibmq\_rochester} device. Out of the 58 qubit pairs,~56 were found to be entangled when using QREM and 31 when not using QREM. The two qubit pairs 13-14 and 12-13 did not have negativity values significantly above zero in either case of QREM or no QREM. When QREM was applied, the resulting values have mean 0.24 with standard deviation 0.12. When QREM was not applied, the mean is~0.06 with standard deviation~0.07. Detected pairs of entanglement are visualized on the device layout in Figure~\ref{fig:entanglement-layouts}a. \textbf{(b)} The negativity results for the  \textit{ibmq\_manhattan} device. All 72 of the qubit pairs were found to be entangled in both cases of QREM and no QREM. When QREM was applied, the resulting values have mean 0.40 with standard deviation 0.05. When QREM was not applied, the values have mean 0.33 with standard deviation 0.05. Detected pairs of entanglement are visualized on the device layout in Figure~\ref{fig:entanglement-layouts}b.} \label{fig:negativity-plots}
\end{figure}

\begin{figure}
	\centering
    \includegraphics[width=0.65\linewidth]{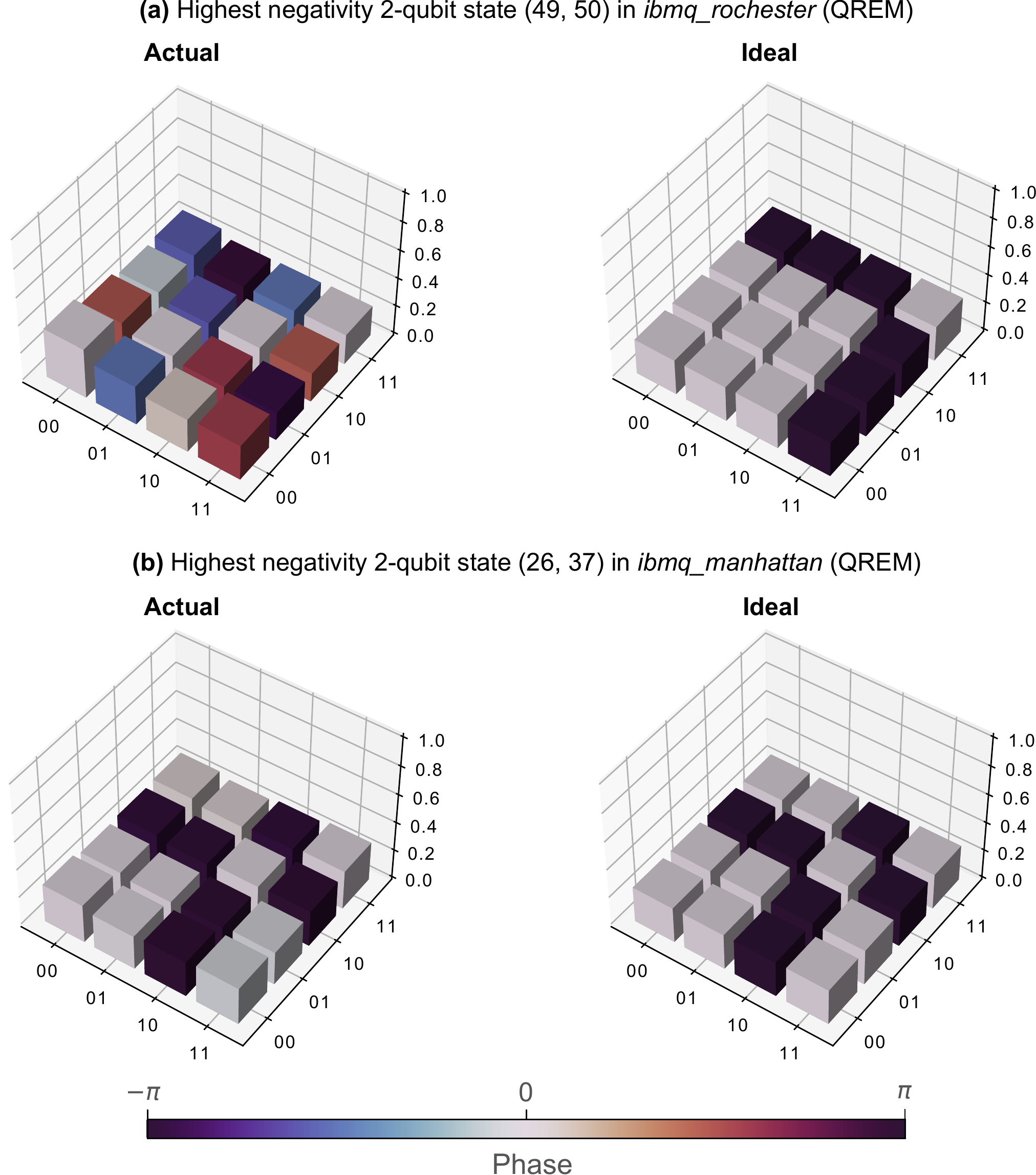}
	\caption{The density matrices for the pairs of qubits found to be most entangled in each device using QREM. When calculating the negativity, the neighboring qubits in the density matrix are projected onto each combination of $Z$-basis states and the combination that produces the highest negativity is chosen to represent the extent of entanglement between the target pair of qubits. These plots show the resulting two-qubit states after projecting the neighboring qubits to $Z$-basis states that maximize the negativity and then partial tracing over them. Ideal states are the corresponding Bell states (up to local transformations) formed by the neighboring qubit projections. \textbf{(a)} The density matrix for qubit pair (49, 50), which had a measured negativity of 0.44, in the \textit{ibmq\_rochester} device after projecting neighboring qubits q41 and q48 onto the 0 state. The density matrices with the highest negativity without QREM and the lowest negativity with and without QREM in the \textit{ibmq\_rochester} device are shown in Figure~\ref{fig:rochester-density-matrices}. \textbf{(b)} The density matrix for qubit pair (26, 37), which had a measured negativity of 0.48, in the \textit{ibmq\_manhattan} device after projecting neighboring qubits q23 and q36 onto the 1 and 0 states respectively. The density matrices with the highest negativity without QREM and the lowest negativity with and without QREM in the \textit{ibmq\_manhattan} device are shown in Figure~\ref{fig:manhattan-density-matrices}} \label{fig:highest-negativity-density-matrices}
\end{figure}

We perform entanglement detection on the now retired 53-qubit \textit{ibmq\_rochester} device which was measured to have a quantum volume of 8~\cite{perelshtein2020advanced} and the 65-qubit \textit{ibmq\_manhattan} device which currently has a quantum volume of 32. Both devices have a heavy-hexagon layout where qubits are located on vertices and edges of a hexagonal grid, as shown in Figure~\ref{fig:entanglement-layouts}. The large numbers of qubits in the devices make them ideal targets for demonstrating the existence of large entangled states. The standard QREM procedure is applied to both devices to help reduce the effects of measurement errors obfuscating entanglement detection. This is especially important for the \textit{ibmq\_rochester} device which has comparatively high qubit readout error rates (an average of 12.6\% compared to 2.1\% for the \textit{ibmq\_manhattan} device during the times that the experimental results were obtained in this work). To measure entanglement, we first prepare a native-graph state containing all qubits and an edge for every connected pair of qubits within the device. The CNOT depth of the preparation circuit for a native-graph state is at least as long as the maximum number of qubits any single qubit is connected to (see Figure~\ref{fig:graph-state-prep-example}). For native-graph states in the \textit{ibmq\_rochester} and \textit{ibmq\_manhattan} devices, the CNOT circuit depths are both three.

For the \textit{ibmq\_rochester} device, we perform tomography on all pairs of qubits and their neighbors using circuits consisting of 4000 shots.
Additionally, there were two calibration circuits, each of 8192 shots, performed for QREM, which was applied to the tomography measurements before constructing density matrices and projecting neighboring qubits onto the $Z$-basis. The calibration circuits were executed separately because the tomography basis measurements are on the prepared graph state, while the calibration measurements are on the combinations of measurement bases in the computational basis (where each qubit is prepared to be either $|0\rangle$ or $|1\rangle$).
Executions of the experiment were completed within the time period between system calibrations, although we found it challenging to do so due to the sheer number of circuits. Some circuits were executed considerably later than others within a calibration period. In particular, for the \textit{ibmq\_rochester} experiments, all tomography circuits were executed before any QREM circuits, which might have reduced the effectiveness of QREM due to the qubit readout error rates drifting over time. However, measurements we conducted on the readout error for each qubit across a single calibration period (17 samples per qubit and 8192 shots per measurement) showed the system to be relatively stable with an average standard deviation of the error rate at the 1\% level. For an analysis of IBM Quantum calibration data in the context of noise modelling, see~\cite{ku2020experimental}. We successfully completed a total of six full executions of the experiments within calibration times. The results for the best of them are shown in Figure~\ref{fig:negativity-plots}a. The figure displays the results for the highest negativity calculated over each combination of $Z$-basis state projections of neighboring qubits. The negativity ranges between 0 and 0.5 where 0 indicates no entanglement and 0.5 indicates maximal entanglement. When QREM was applied, full 53-qubit entanglement was detected within the system with 56 of the 58 pairs shown to be above zero negativity with 95\% confidence, visualized in Figure~\ref{fig:entanglement-layouts}a. The negativity values with QREM applied had a mean of 0.24 with standard deviation 0.12. When QREM was not applied,~31 of the~58 pairs were found to be significantly above zero negativity with the largest entangled region consisting of~9 qubits. The negativity values in this case had a mean of 0.06 with standard deviation 0.07. The density matrix for the qubit pair exhibiting the highest negativity with QREM is shown in Figure~\ref{fig:highest-negativity-density-matrices}. The density matrices for the highest negativity pair without QREM and the lowest negativity pair with and without QREM are shown in Figure~\ref{fig:rochester-density-matrices}. The other five experiments on \textit{ibmq\_rochester} that ran to completion demonstrated entangled regions of 53, 52, 52, 38, and 21 qubits when QREM was applied. 

For the \textit{ibmq\_manhattan} device, tomography was performed with the same approach as the \textit{ibmq\_rochester} device. Although, to include device drift noise in the error bars when applying QREM, the calibration data was not bootstrapped from a single source of measured data. Instead, we gathered data from~50 calibration experiments across the execution period and applied QREM from each of them to~50 copies of the tomography data, which were then each resampled to become the bootstrapped samples. The calibration experiments were performed such that~25 were executed before the tomography circuits and~25 were executed afterwards to help account for the difference in readout errors of the device across the execution period. We completed one full execution of the experiment between device calibration times with results shown in Figure~\ref{fig:negativity-plots}b. We found that all 72 of the connected pairs of qubits had negativity above zero with 95\% confidence in both cases of applying and not applying QREM. Thus the entangled region of the device clearly included all 65 qubits. The negativity values when QREM was applied had a mean of 0.40 with standard deviation 0.05, and when QREM was not applied, had a mean of 0.33 with standard deviation 0.05. The density matrix for the qubit pair exhibiting the highest negativity with QREM is shown in Figure~\ref{fig:highest-negativity-density-matrices}. The density matrices for the highest negativity pair without QREM and the lowest negativity pair with and without QREM are shown in Figure~\ref{fig:manhattan-density-matrices}. 

During the calibration periods for these experiments, the average qubit readout error rate for the \textit{ibmq\_rochester} device was 12.6\% with a standard deviation of 9.3\% across the qubits and the average CNOT error rate was 4.6\% with a standard deviation of 2.4\%. For the \textit{ibmq\_manhattan} device, the average qubit readout error rate was 2.1\% with a standard deviation of 1.5\% across the qubits and the average CNOT error rate was 1.5\% with a standard deviation of 0.6\%. More detail can be found in Appendix~\ref{sec:device-specifications} and Figure~\ref{fig:error-maps}. Although the results of this work show full bipartite entanglement across each of the two devices, they are inconclusive as to what level of entanglement is present, for example whether the states are genuinely multipartite entangled. 

In previous work~\cite{mooney2019entanglement}, graph state entanglement witnesses~\cite{toth2005entanglement, zhou2019detecting} were measured across a prepared graph state over the full 20-qubit \textit{ibmq\_poughkeepsie} device, to detect the presence of genuine multipartite entanglement within regions of the state. Entanglement witness protocols typically provide methods to measure sufficient but not necessary conditions for genuine multipartite entanglement. Due to the noise within the \textit{ibmq\_poughkeepsie} device, genuine multipartite entanglement was only detected on up to 3-qubit regions within the prepared state. It would be interesting to investigate this question thoroughly for graph states in future research, especially with the application of QREM.

The confidence intervals for all negativity values in our data are calculated using bootstrapping with bias correction~\cite{efron1986bootstrap, efron1981nonparametric}. Without bias correction, we found that the negativity sample estimates were often above the bootstrapped sample means and in some cases were above the bootstrapped confidence intervals entirely. This suggests that the bootstrapped negativity is a biased estimate of the sample negativity when using 4000 shots in the experiments, thus bias correction is used to adjust the confidence intervals to match the negativity estimate. This also suggests that the negativity estimate is biased with respect to the population negativity, thus we could further bias correct the estimate and the confidence intervals upwards again to approximate the population. However, this assumes that the bias between the bootstrap and the sample is the same as the bias between the sample and the population. This extra step is not required to show entanglement within our experiments so we do not use it here. To perform bootstrapping on our data, the data are partitioned into tomography measurement related data and QREM calibration related data. For each bootstrap sample, a copy of the tomography data is made. The QREM data are bootstrapped by resampling with replacement, then applied to each of the copies of tomography data, which are then each resampled once to produce the bootstrapped QREM-applied tomography samples. The negativities are then calculated for each sample to produce the final bootstrapped distribution of negativities which are used to calculate the 95\% confidence intervals. For pairs of connected qubits with low values of negativity where some of the bootstrapped values are zero, bias correction leaves the bottom of the distribution clamped disallowing bootstrapped values lower than the estimate subtracted by the bootstrap sample mean. This could be a problem, for example, if all bootstrapped values for the lower bound of the confidence interval are zero, and are bias corrected to be equal to the mean, which may be above zero, leading to the full confidence interval being above zero. To help avoid this problem, the zero values are ignored when calculating the mean of the bootstrap sample negativities, which increases the mean and consequently decreasing the amount of bias correction applied to the bootstrapped sample negativities upwards, leading to a more robust error bar. Applying bias correction did not change whether any of the confidence intervals include the zero negativity value or not, hence the conclusion of these results is unaffected.

\section{Discussion}

We prepared native-graph states on the IBM Quantum \textit{ibmq\_rochester} and \textit{ibmq\_manhattan} devices and measured the level of two-qubit entanglement across the states. Full quantum state tomography was performed on each connected pair of qubits and their neighbors within the graph states. The neighboring qubits for each of the pairs were projected onto the $Z$-basis and the negativities were measured from the resulting density matrix. To reduce the effects of individual qubit noise introduced during the measurement process, we performed the experiments with and without QREM. For the now retired \textit{ibmq\_rochester} device which had a quantum volume of 8, a total of 56 out of 58 of the connected pairs of qubits were found to be entangled when QREM was applied, resulting in the region of entangled qubits spanning all 53 qubits of the device, with a mean negativity of 0.24 and a standard deviation of 0.12. When QREM was not used, 31 out of the 58 connected pairs had a negativity significantly above zero with 95\% confidence, resulting in the largest region of entangled qubits consisting of 9 qubits. This signifies the power of QREM in devices with relatively high readout error rates such as the \textit{ibmq\_rochester}.
For the \textit{ibmq\_manhattan} device which at the time of these experiments had a quantum volume of 32, all 72 connected pairs of qubits were found to be entangled in both cases of QREM and no QREM, with a mean negativity of 0.40 and a standard deviation of 0.05 when QREM was applied. This indicates that the region of entangled qubits in both cases of QREM and no QREM included all qubits of the device.
The results of this work demonstrate the capability of whole-device bipartite entanglement within two of the largest IBM Quantum superconducting devices to date and indicate the positive direction towards the physical computation of sizeable and complex quantum algorithms.

\medskip
\textbf{Acknowledgments} \par 
This work was supported by the University of Melbourne through the establishment of an IBM Quantum Network Hub at the University. CDH is supported by a research grant from the Laby Foundation. We would like to thank Howard Bondell for valuable discussions on aspects of the statistical analysis.

\medskip
\textbf{Conflict of Interest}
\emph{Non-financial competing interests}: The authors are supported by the University of Melbourne through the establishment of an IBM Quantum Network Hub at the University.

\medskip
\textbf{Data Availability Statement}
The datasets generated and/or analyzed during the current study are available from the corresponding author on reasonable request.

\medskip

%
\bibliographystyle{unsrtnat}
\bibliography{}

\section*{Appendix}
\appendix
\renewcommand\thefigure{\thesection.\arabic{figure}}
\setcounter{figure}{0} 
\renewcommand\thedefinition{\thesection.\arabic{definition}}
\setcounter{definition}{0} 
\renewcommand\thelemma{\thesection.\arabic{lemma}}
\setcounter{lemma}{0}
\renewcommand\thetheorem{\thesection.\arabic{theorem}}
\setcounter{theorem}{0}
\renewcommand\thecorollary{\thesection.\arabic{corollary}}
\setcounter{corollary}{0}

\section{Reduced density matrices for connected pairs of qubits}\label{sec:appendix-dm}

\begin{figure}
	\centering
    \includegraphics[width=0.65\linewidth]{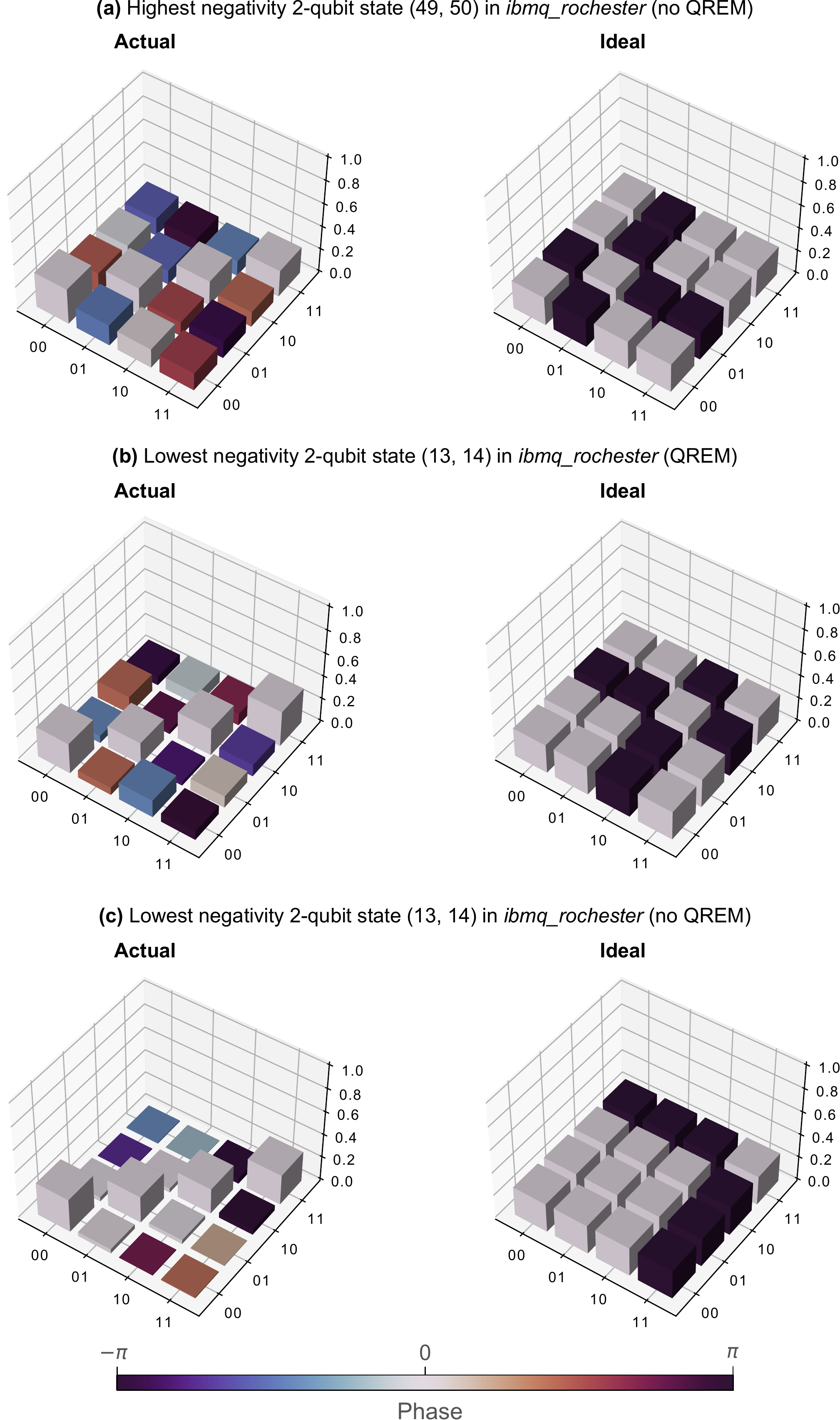}
	\caption{The density matrices for the highest and lowest negativity pairs of qubits in the \textit{ibmq\_rochester} device. The highest negativity pair with QREM is shown in Figure~\ref{fig:highest-negativity-density-matrices}. These plots show the resulting two-qubit states after projecting the neighboring qubits to $Z$-basis states that maximize the negativity and then partial tracing over them. Ideal states are the corresponding Bell states (up to local transformations) formed by the neighboring qubit projections. \textbf{(a)} The qubit pair (49, 50) without QREM, which had a measured negativity of 0.21 after projecting neighboring qubits q41 and q48 to the 1 and 0 states respectively. \textbf{(b)} The qubit pair (13, 14) with QREM, which had a measured negativity of 0 after projecting qubits q6, q12 and q15 to the 1, 0 and 0 states respectively. \textbf{(c)} The qubit pair (13, 14) without QREM, which had a measured negativity of 0 after projecting qubits q6, q12 and q15 to the 0 state.} \label{fig:rochester-density-matrices}
\end{figure}

\begin{figure}
	\centering
    \includegraphics[width=0.65\linewidth]{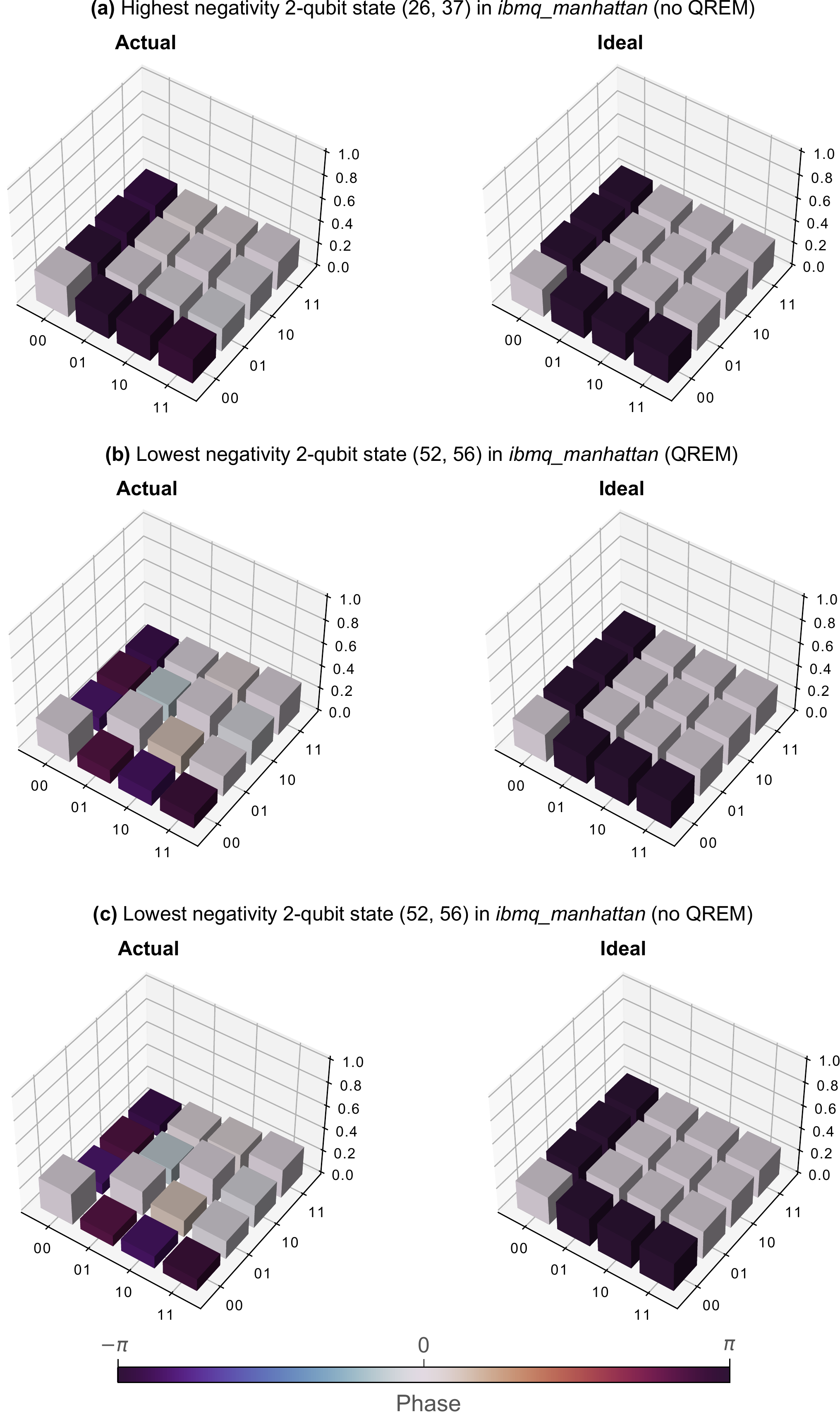}
	\caption{The density matrices for the highest and lowest negativity pairs of qubits in the \textit{ibmq\_manhattan} device. The highest negativity pair with QREM is shown in Figure~\ref{fig:highest-negativity-density-matrices}. These plots show the resulting two-qubit states after projecting the neighboring qubits to $Z$-basis states that maximize the negativity and then partial tracing over them. Ideal states are the corresponding Bell states (up to local transformations) formed by the neighboring qubit projections. \textbf{(a)} The qubit pair (26, 37) without QREM, which had a measured negativity of 0.41 after projecting neighboring qubits q23 and q36 to the 1 state. \textbf{(b)} The qubit pair (52, 56) with QREM, which had a measured negativity of 0.23 after projecting qubits q43, q55 and q57 to the 1, 1 and 0 states respectively. \textbf{(c)} The qubit pair (52, 56) without QREM, which had a measured negativity of 0.17 after projecting qubits q43, q55 and q57 to the 1, 1 and 0 states respectively.} \label{fig:manhattan-density-matrices}
\end{figure}

After performing tomography on a connected pair of qubits and their neighbors and projecting the neighboring qubits to $Z$-basis states, the pair ideally forms a Bell pair up to local transformations. To help conceptually associate the resulting density matrices and the corresponding negativities, the density matrices for the highest and lowest negativity qubit pairs for the cases of QREM and no QREM on both devices were plotted along with the ideal density matrices. The highest negativity pairs with QREM on both devices are shown in Figure~\ref{fig:highest-negativity-density-matrices}. For the \textit{ibmq\_rochester} device, density matrices for the highest negativity pair without QREM and the lowest negativity pair with and without QREM are shown in Figure~\ref{fig:rochester-density-matrices}. The equivalent density matrices for the \textit{ibmq\_manhattan} device are shown in Figure~\ref{fig:manhattan-density-matrices}.

\renewcommand\thefigure{\thesection.\arabic{figure}}
\setcounter{figure}{0} 
\renewcommand\thedefinition{\thesection.\arabic{definition}}
\setcounter{definition}{0} 
\renewcommand\thelemma{\thesection.\arabic{lemma}}
\setcounter{lemma}{0}
\renewcommand\thetheorem{\thesection.\arabic{theorem}}
\setcounter{theorem}{0}
\renewcommand\thecorollary{\thesection.\arabic{corollary}}
\setcounter{corollary}{0}
\section{Device calibration data}\label{sec:device-specifications}

\begin{figure}
	\centering
	\includegraphics[scale=1]{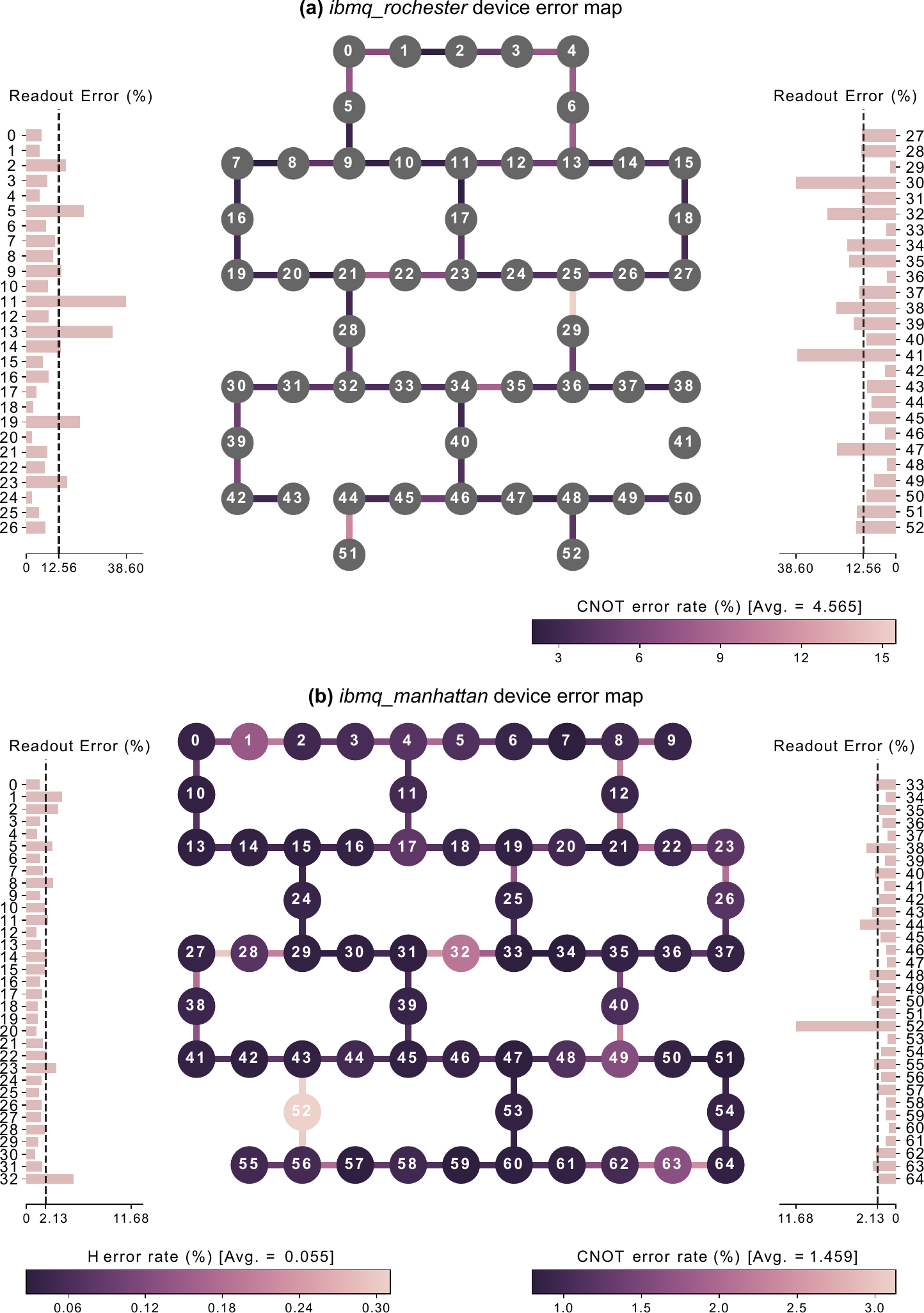}
	\caption{The error maps corresponding to the calibration periods of the experiments performed in this work. These were obtained using code from the IBM Quantum Qiskit python framework~\cite{Qiskit}. The color legends on the nodes and edges correspond to Hadamard gate and CNOT error rates respectively (for the \textit{ibmq\_rochester} device, CNOTs only). \textbf{(a)} The \textit{ibmq\_rochester} device. Three edges were removed from the map because the corresponding CNOT error rates were undefined in our data. These edges were (38, 41), (41, 50) and (43, 44). \textbf{(b)} The \textit{ibmq\_manhattan} device. } \label{fig:error-maps}
\end{figure}

Due to temporal instability of the internal control parameters and external influences on quantum devices, the parameters of the devices drift over time~\cite{proctor2020detecting}. To alleviate this, IBM Quantum devices undergo routine calibrations. Each calibration produces different error rates in the device. For the \textit{ibmq\_rochester} device, during the calibration period for which the results in this work were obtained, the average qubit readout error rate was 12.6\% with a standard deviation of 9.3\% and the average CNOT error rate was 4.6\% with a standard deviation of 2.4\%. For the \textit{ibmq\_manhattan} device, during its calibration period, the average qubit readout error rate was 2.1\% with a standard deviation of 1.5\% and the average CNOT error rate was 1.5\% with a standard deviation of 0.6\%. The error maps for these experiment calibration periods were constructed using the IBM Quantum Qiskit python framework~\cite{Qiskit} and are displayed in Figure~\ref{fig:error-maps}.

\end{document}